\documentclass[conference]{IEEEtran}

\usepackage[utf8]{inputenc}
\usepackage[T1]{fontenc}
\usepackage{cite}
\usepackage{amsmath,amssymb}
\usepackage{graphicx}
\graphicspath{{images/}}
\usepackage[caption=false,font=footnotesize]{subfig}
\usepackage{array}
\usepackage{booktabs}
\usepackage{url}
\usepackage{tabularx}
\usepackage{ragged2e}
\usepackage{microtype}
\usepackage{eso-pic} 

\usepackage{balance}
\newcommand{\psse}{PSS\textsuperscript{\textregistered}E}
\newcommand{\regca}{REGCA}
\newcommand{\reecc}{REECC}
\newcommand{\repca}{REPCA}
\newcolumntype{Y}{>{\RaggedRight\arraybackslash}X}

\begin{document}

\AddToShipoutPictureFG*{%
  \AtPageLowerLeft{%
    \put(36,30){%
      \parbox{540pt}{\centering\fontsize{7}{8.5}\selectfont\itshape
      Accepted author manuscript for presentation at IEEE PES Grid Edge Technologies 2027,\
      April 19--22, 2027, Salt Lake City, Utah, USA.\par
      This accepted version may differ from the final version published in IEEE Xplore.
      }%
    }%
  }%
}

\title{Control-Layer Interpretation of Reactive Power Recovery Limits in EMT--RMS Model Benchmarking of Inverter-Based Resources}

\author{
\IEEEauthorblockN{Dileep Cheerla, Amirthagunaraj Yogarathnam, Divya Jain, and Satish Natti}
\IEEEauthorblockA{GridAxon, Inc., USA\\
Email: \{dileep, araj, divya, \& satish\}@gridaxon.io}
}

\maketitle

\begin{abstract}
EMT--RMS benchmarking is widely used to verify cross-platform consistency of inverter-based resource (IBR) dynamic models. This paper examines a 200~MW anonymized IBR represented in PSCAD and \psse{}. The EMT model separates inverter electrical capability from a plant-level power plant controller (PPC) recovery target, whereas the RMS model represents these functions through \reecc{} and \repca{} parameters. Directly assigning the PPC target of $\pm0.2764$~pu to \repca{} hard reactive limits is initialization-inconsistent at the original lagging operating point because the initial reactive command is 0.396273~pu. A second, near-unity-power-factor operating point initializes successfully with the same diagnostic limit, demonstrating that the conflict is operating-point dependent. Using the non-conflicting capability mapping removes the original hard-limit conflict, but the benchmark still shows 18.5--30.2~MVAr post-event POI reactive-power differences from PSCAD. A one-at-a-time \repca{} $K_c$ sensitivity under matched benchmark initial conditions further shows that RMS recovery is parameter- and limit-dependent: at $K_c=0.05$ and 0.10 the aggregate machine reactive output reaches the configured capability bounds, while $K_c=0.15$ moves VDOWN, VUP, and HVRT off those bounds and changes their settled POI reactive power by 8.94, 18.00, and 3.17~MVAr, respectively. Machine-to-POI accounting further shows why machine-side saturation need not reproduce the POI-level PPC recovery magnitude. The results distinguish initialization feasibility from recovery equivalence and provide a control-layer diagnostic workflow beyond waveform comparison.
\end{abstract}


\section{Introduction}
Electromagnetic transient (EMT) and positive-sequence root-mean-square (RMS) models are widely used in IBR interconnection studies, including grid-edge distributed energy resources, battery energy storage systems, renewable generation, and large flexible-load interconnections. Recent verification and co-simulation workflows emphasize EMT--RMS consistency because generic positive-sequence models reduce detailed converter, protection, and plant-controller behavior~\cite{nerc_emt,wecc_posseq,pjm_emt_guidelines,wang_cosim_2025}. Related studies have also examined model adequacy and aggregation effects under transient conditions~\cite{Raj_GFC_adeq,Raj_DERAggrigat}. Complementary work has examined fault-resilient grid-forming converter control and auxiliary battery-energy-storage control for LVRT and load-smoothing applications, providing related examples of converter- and plant-level control behavior in modern IBR applications~\cite{dileep1,dileep2}.

Waveform overlay is an important first benchmarking check, but agreement or disagreement alone does not identify whether the underlying control quantities have equivalent physical meaning. A reactive quantity in an EMT model may represent converter electrical capability, whereas a numerically similar quantity in an RMS plant controller may act as an operational constraint. WECC guidance distinguishes plant-controller quantities from converter-level controls and emphasizes that generic positive-sequence models remain reduced-order representations whose applicability requires engineering judgment~\cite{wecc_posseq,wecc_repc}.

Building on verification and model-quality workflows that use prescribed disturbance-response checks~\cite{nerc_emt,ercot_mqt}, the methodological contribution of this paper is a compact three-stage diagnostic workflow that extends waveform comparison by: (i) identifying the control-layer meaning and common per-unit base of reactive parameters; (ii) testing hard-limit initialization feasibility across operating points; and (iii) separating hard-limit effects from plant-controller parameter effects using controlled sensitivity. The distinction is that parameter meaning, initialization feasibility, and recovery sensitivity are evaluated as separate checks rather than treating waveform agreement as the sole diagnostic. The study uses one anonymized 200~MW plant and therefore does not claim statistical generality across vendors, grid strengths, or control structures. Instead, the numerical results illustrate a repeatable diagnostic sequence for EMT models that separate electrical capability from PPC recovery logic and RMS models that represent these functions through generic control blocks.

\section{Benchmark System and Control Mapping}
\begin{figure}[!t]
\centering
\includegraphics[width=0.84\linewidth]{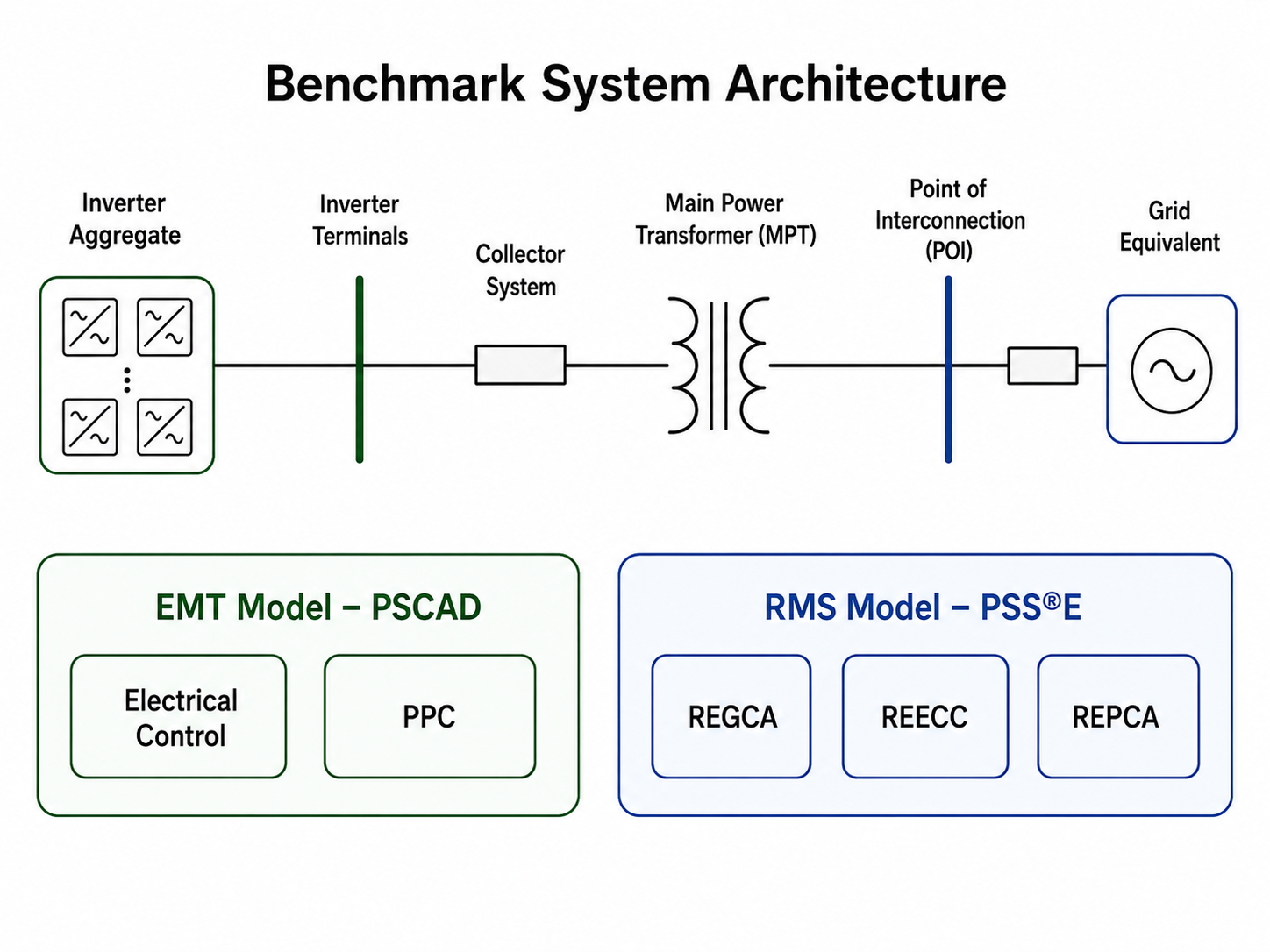}
\caption{Simplified benchmark architecture and control-layer representation.}
\label{fig:architecture}
\end{figure}

The plant is connected to the point of interconnection (POI) through an aggregated collector system and main power transformer. Fig.~\ref{fig:architecture} shows the benchmark architecture, and Table~\ref{tab:case_summary} summarizes the principal benchmark quantities. PSCAD~\cite{pscad} contains separate inverter electrical-control and PPC layers, while the \psse{} model uses the generic \regca{}, \reecc{}, and \repca{} blocks~\cite{psse_manual,epri_model_guide,wecc_models,wecc_pv_guideline}. WECC's generic renewable-energy-system guidance documents the modular generator/converter, electrical-control, and plant-controller model structure and its implementation across major positive-sequence simulation platforms~\cite{wecc_models}.

\begin{table}[!t]
\caption{Benchmark Case Summary}
\label{tab:case_summary}
\centering
\scriptsize
\setlength{\tabcolsep}{2pt}
\begin{tabularx}{\columnwidth}{|p{0.39\columnwidth}|Y|}
\hline
\textbf{Parameter} & \textbf{Value} \\
\hline
Plant POI rating & 200~MW \\
\hline
Aggregate inverter terminal rating & 277.2~MVA \\
\hline
PPC / RMS control base & 238.072~MVA \\
\hline
Reactive recovery basis & 0.95 power factor at POI \\
\hline
Studied PPC recovery target & 65.8~MVAr (0.2764~pu) \\
\hline
Electrical capability & $Q_{\max}=0.504889$~pu, $Q_{\min}=-0.287728$~pu \\
\hline
Disturbances & VDOWN, VUP, HVRT, LVRT \\
\hline
\end{tabularx}
\end{table}

The 0.95 power-factor recovery basis is consistent with the reactive-power framework in FERC Order No.~827 and broader IBR performance requirements in IEEE Std 2800-2022~\cite{ferc827,ieee2800}; related steady-state work has separately examined POI reactive deliverability and equipment capability~\cite{cheerla_naps_2026}. For this study, however, the PPC value is a supervisory recovery target rather than the full inverter capability. A 0.95-power-factor basis at 200~MW corresponds to approximately 65.74~MVAr, while the studied PPC target is represented as 0.2764~pu on the 238.072~MVA control base, corresponding to approximately 65.80~MVAr:
\begin{equation}
\begin{aligned}
Q_{0.95}&=200\tan(\cos^{-1}0.95)\approx65.74~\mathrm{MVAr},\\
Q_{\mathrm{target}}&=0.2764\times238.072\approx65.80~\mathrm{MVAr}.
\end{aligned}
\label{eq:qtarget}
\end{equation}

Two RMS mappings are evaluated in Table~\ref{tab:mappings}. In both mappings, the \reecc{} limits retain the electrical-control capability values. In the capability-based baseline, the \repca{} bounds are set to the same wider numerical values so that the plant-controller output bounds do not conflict with the intended initial operating point. In the diagnostic mapping, only the \repca{} bounds are narrowed to the PPC recovery target of $\pm0.2764$~pu to test whether direct transfer of a supervisory target into a hard plant-controller bound preserves initialization and recovery behavior. This diagnostic mapping is not proposed as a production representation. This interpretation is consistent with guidance that \repca{} output limits must be interpreted according to the active plant-control structure rather than as stand-alone inverter capability~\cite{wecc_repc,epri_model_guide}.

\begin{table}[!t]
\caption{RMS Reactive-Limit Mappings Evaluated}
\label{tab:mappings}
\centering
\scriptsize
\setlength{\tabcolsep}{1.5pt}
\begin{tabular}{|c|c|c|}
\hline
\textbf{Mapping} & \textbf{\reecc{} $Q_{\max}/Q_{\min}$} & \textbf{\repca{} $Q_{\max}/Q_{\min}$} \\
\hline
Capability & $+0.504889/-0.287728$ & $+0.504889/-0.287728$ \\
\hline
Diagnostic & $+0.504889/-0.287728$ & $+0.2764/-0.2764$ \\
\hline
\end{tabular}
\par\scriptsize Values are per unit on the 238.072~MVA control base.
\end{table}

\section{Initialization and Operating-Point Dependence}
For the reactive-command path considered here, initialization consistency requires
\begin{equation}
Q_{\min}^{\mathrm{hard}}\le Q_{0}^{\mathrm{cmd}}\le Q_{\max}^{\mathrm{hard}}.
\label{eq:init}
\end{equation}
At the original lagging operating point, $Q_{0}^{\mathrm{cmd}}=0.396273$~pu exceeds the diagnostic upper limit of 0.276400~pu, and \psse{} reports an initial-condition suspect. This is a mapping inconsistency, not a simulator deficiency.

To test operating-point dependence, the same diagnostic limits were retained while the power-flow operating point was changed to near-unity power factor at the POI. The grid-side reactive exchange decreased in magnitude from approximately 66.1 to 0.4~MVAr, while the aggregate machine reactive output decreased from 94.34 to 25.93~MVAr; the \psse{} initial-condition check then passed. Table~\ref{tab:init_sensitivity} therefore demonstrates that the initialization conflict is operating-point dependent and is not inherent to the model at every operating point.

\begin{table}[!t]
\caption{Initialization Check at Two Operating Points}
\label{tab:init_sensitivity}
\centering
\scriptsize
\setlength{\tabcolsep}{1.8pt}
\begin{tabularx}{\columnwidth}{|p{0.25\columnwidth}|p{0.20\columnwidth}|p{0.22\columnwidth}|Y|}
\hline
\textbf{Operating Point} & \textbf{Grid-side $|Q|$} & \textbf{Machine Q} & \textbf{Result} \\
\hline
Original lagging & $\approx66.1$~MVAr & 94.34~MVAr & Suspect \\
\hline
Near-unity PF & $\approx0.4$~MVAr & 25.93~MVAr & Check O.K. \\
\hline
\end{tabularx}
\end{table}

\section{Disturbance Profiles and Evaluation Window}
The voltage-disturbance stimuli are summarized in Table~\ref{tab:profiles}. They are applied consistently for EMT--RMS comparison and are not presented as a complete ride-through compliance demonstration. ERCOT model-quality-test guidance includes voltage-step and ride-through tests in dynamic-model evaluation workflows~\cite{ercot_mqt}. The quantitative comparison uses the 12.5--20~s post-event window after the dominant response has completed.

\begin{table}[!t]
\caption{Applied Voltage Disturbance Profiles}
\label{tab:profiles}
\centering
\scriptsize
\setlength{\tabcolsep}{2pt}
\begin{tabularx}{\columnwidth}{|p{0.14\columnwidth}|Y|}
\hline
\textbf{Case} & \textbf{Voltage Profile} \\
\hline
VDOWN & 1.00~pu through 5.0~s; step to 0.97~pu at 5.0~s \\
\hline
VUP & 1.00~pu through 5.0~s; step to 1.03~pu at 5.0~s \\
\hline
HVRT & 1.00~pu through 8.0~s; 1.20~pu (8.0--8.2~s); 1.175~pu (8.2--8.5~s); 1.15~pu (8.5--9.0~s); then 1.10~pu \\
\hline
LVRT & 1.00~pu through 8.0~s; 0~pu (8.0--8.15~s); recovery ramp to 0.90~pu by 9.75~s \\
\hline
\end{tabularx}
\end{table}

\section{Benchmark Results}
PSCAD--EMT denotes the detailed EMT benchmark, \psse{}--RMS denotes the capability-based mapping, and \psse{}--RMS* denotes the diagnostic mapping. PSCAD is treated as the most detailed available integrated study representation, not as field validation or absolute plant truth; NERC guidance distinguishes model verification from validation against actual equipment~\cite{nerc_emt}. The active-power responses remain closely aligned for the studied disturbances; therefore, the discussion focuses on the reactive-power recovery differences.

\begin{figure*}[!t]
\vspace{-15pt}
\centering
\subfloat[VDOWN\label{fig:vdown_threeway}]{
\includegraphics[width=0.4\textwidth]{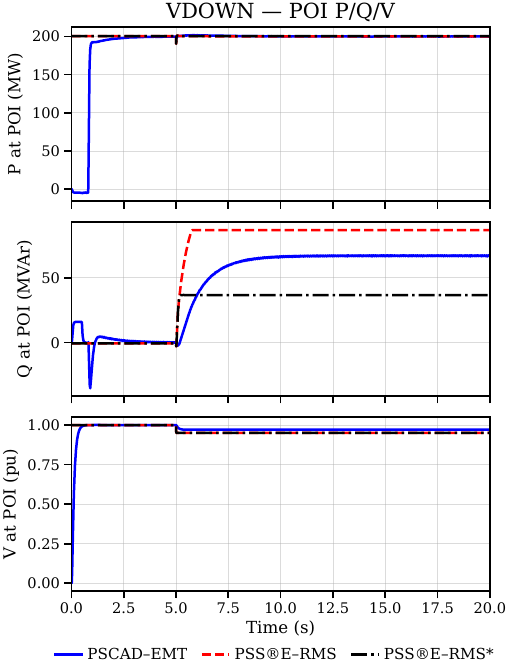}}
\hfill
\subfloat[VUP\label{fig:vup_threeway}]{
\includegraphics[width=0.4\textwidth]{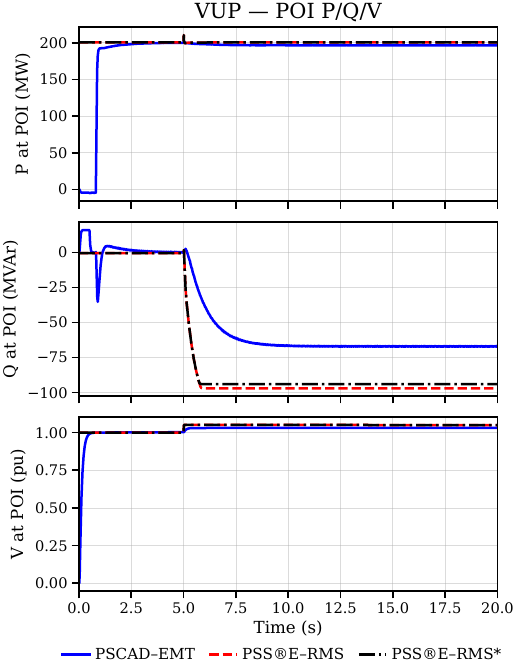}}
\caption{POI active power, reactive power, and voltage comparison for VDOWN and VUP.}
\label{fig:voltage_step_results}
\end{figure*}

\begin{figure*}[!t]
\vspace{-10pt}
\centering
\subfloat[HVRT\label{fig:hvrt_threeway}]{
\includegraphics[width=0.4\textwidth]{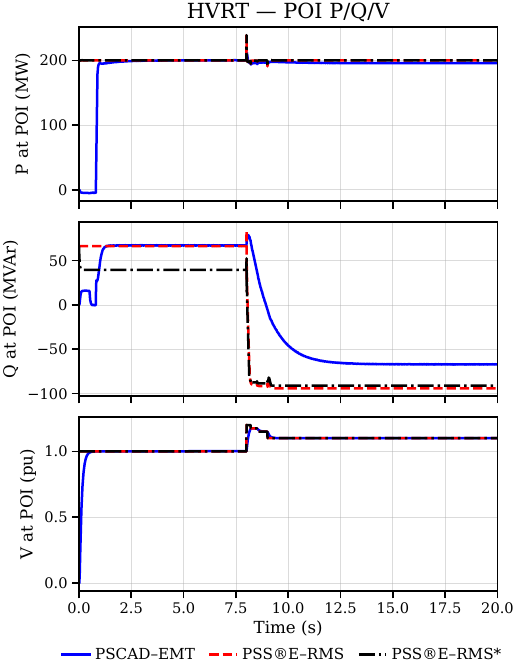}}
\hfill
\subfloat[LVRT\label{fig:lvrt_threeway}]{
\includegraphics[width=0.4\textwidth]{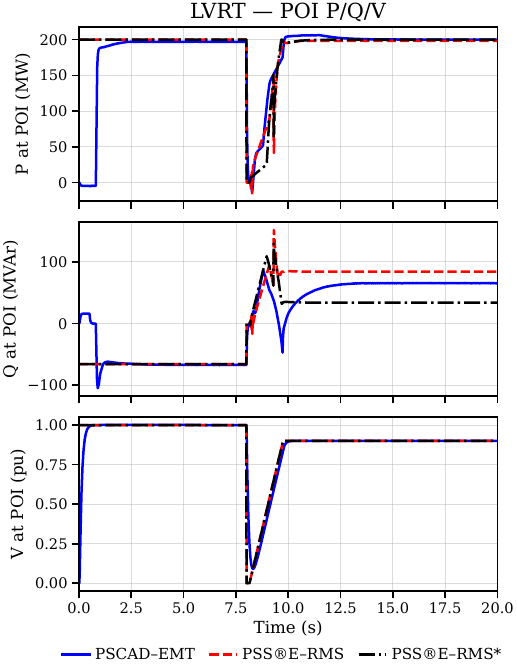}}
\caption{POI active power, reactive power, and voltage comparison for HVRT and LVRT.}
\label{fig:vrt_results}
\end{figure*}

Table~\ref{tab:init_sensitivity} establishes that the diagnostic initialization conflict is operating-point dependent. With the non-conflicting capability mapping, Figs.~\ref{fig:voltage_step_results} and~\ref{fig:vrt_results} show that valid initialization alone does not reproduce the EMT post-event reactive recovery. Table~\ref{tab:q_summary} quantifies the benchmark: the capability-based RMS case differs from PSCAD by 18.53--30.18~MVAr. For the diagnostic mapping, VDOWN, VUP, and leading LVRT initialize successfully yet differ from EMT by 27.16--31.76~MVAr, while the lagging-HVRT diagnostic case is initialization-suspect and its post-event value is retained only as diagnostic evidence. Thus, direct hard-limit mapping can distort recovery even when initialization itself is valid.

\begin{table}[!t]
\caption{Post-Event POI Q Comparison Over 12.5--20~s}
\label{tab:q_summary}
\centering
\scriptsize
\setlength{\tabcolsep}{1.25pt}
\begin{tabular}{|c|r|r|r|r|r|}
\hline
\textbf{Case} & \textbf{EMT} & \textbf{RMS} & \textbf{RMS*} & \textbf{$|\Delta|$} & \textbf{$|\Delta^*|$} \\
\hline
VDOWN & 66.99 & 86.54 & 36.43 & 19.55 & 30.56 \\
\hline
VUP & $-67.08$ & $-97.26$ & $-94.24$ & 30.18 & 27.16 \\
\hline
HVRT (lag) & $-66.83$ & $-94.21$ & $-91.21$ & 27.38 & 24.38 \\
\hline
LVRT (lead) & 65.12 & 83.65 & 33.36 & 18.53 & 31.76 \\
\hline
\end{tabular}
\par\scriptsize All values are MVAr; $|\Delta|=|\mathrm{RMS}-\mathrm{EMT}|$ and $|\Delta^*|=|\mathrm{RMS^*}-\mathrm{EMT}|$. RMS* initialization is O.K. for VDOWN, VUP, and LVRT (lead), and suspect for HVRT (lag).
\end{table}

\section{Plant-Controller Sensitivity and Reactive-Limit Diagnosis}
The baseline comparison establishes that removing the initialization conflict does not, by itself, establish post-event recovery equivalence. To provide quantitative evidence on the influence of the RMS plant-controller representation, a one-factor-at-a-time sensitivity was performed on \repca{} parameter $K_c$, which participates in the plant-level voltage-control path of the generic REPC\_A representation~\cite{epri_model_guide}. $K_c$ was selected because it is active in the configured plant-level voltage/reactive-control path and can be varied independently without changing the selected control mode or the reactive-limit mapping. The nominal value is $K_c=0.10$; only $K_c$ was changed to 0.05 and 0.15, while the remaining RMS parameters and the wider \reecc{}/\repca{} bounds were held fixed.

For each disturbance family, the same pre-event operating condition is retained for all three $K_c$ values. The VDOWN and VUP step cases use a separate near-unity-power-factor power-flow solution from the initialization check in Table~\ref{tab:init_sensitivity} and begin at approximately 200.39~MW and $-0.75$~MVAr at the POI. The HVRT case begins at approximately 0.95 lagging power factor (199.99~MW and $+66.08$~MVAr), and the LVRT case begins at approximately 0.95 leading power factor (199.99~MW and $-66.13$~MVAr). Thus, $K_c$ is varied without changing the initial operating point associated with a given benchmark disturbance. Table~\ref{tab:kc_sensitivity} reports the resulting settled POI reactive-power values.

\begin{table}[!t]
\caption{REPCA $K_c$ Sensitivity Under Matched Benchmark Initial Conditions}
\label{tab:kc_sensitivity}
\centering
\scriptsize
\setlength{\tabcolsep}{2.2pt}
\begin{tabular}{|c|r|r|r|}
\hline
\textbf{Case} & \textbf{$K_c=0.05$} & \textbf{$K_c=0.10$} & \textbf{$K_c=0.15$} \\
\hline
VDOWN & 86.54 & 86.54 & 77.60 \\
\hline
VUP & $-97.26$ & $-97.26$ & $-79.26$ \\
\hline
HVRT (lag) & $-94.21$ & $-94.21$ & $-91.04$ \\
\hline
LVRT (lead) & 83.65 & 83.65 & 83.65 \\
\hline
\end{tabular}
\par\scriptsize Settled POI reactive power over 12.5--20~s, in MVAr.
\end{table}

The $K_c=0.05$ and 0.10 cases are identical to the reported precision because the aggregate machine-terminal reactive output reaches the common configured capability bound in all four disturbances: $+120.20$~MVAr for VDOWN and leading LVRT, and $-68.50$~MVAr for VUP and lagging HVRT. These values correspond to the limits in Table~\ref{tab:mappings} on the 238.072~MVA control base. At $K_c=0.15$, machine reactive output settles at $+110.25$, $-52.37$, $-65.62$, and $+120.20$~MVAr for VDOWN, VUP, lagging HVRT, and leading LVRT, respectively. Accordingly, VDOWN, VUP, and HVRT move off the bound and their settled POI reactive power shifts by 8.94, 18.00, and 3.17~MVAr, while LVRT remains at the positive bound and is unchanged to the reported precision. Because \reecc{} and \repca{} use the same capability values in this mapping, the result identifies saturation in the configured RMS reactive-control path but does not uniquely assign the active limiter to one block.

The machine-to-POI reactive-power accounting further explains the settled mismatch. Define the net reactive offset of the aggregated collector, transformer, and network representation as $\Delta Q_{\mathrm{net}}=Q_{\mathrm{mach}}-Q_{\mathrm{POI}}$. When the RMS response is saturated, $Q_{\mathrm{mach}}\approx Q_{\mathrm{bound}}$, so the settled POI response follows
\begin{equation}
Q_{\mathrm{POI,RMS}}\approx Q_{\mathrm{bound}}-\Delta Q_{\mathrm{net}}.
\label{eq:q_accounting}
\end{equation}
For the nominal $K_c=0.10$ capability cases, the observed $\Delta Q_{\mathrm{net}}$ values are approximately 33.66, 28.76, 25.71, and 36.55~MVAr for VDOWN, VUP, lagging HVRT, and leading LVRT, respectively. For example, the VDOWN relation $120.20-33.66=86.54$~MVAr reproduces the settled RMS value in Table~\ref{tab:q_summary}; relative to the EMT value of 66.99~MVAr, the resulting mismatch is 19.55~MVAr. As a cross-check, offsets computed from the off-bound $K_c=0.15$ runs for VDOWN, VUP, and lagging HVRT differ from the nominal values by only 1.01, 1.87, and 0.29~MVAr; using those offsets in~\eqref{eq:q_accounting} reproduces the saturated $K_c=0.10$ POI values within 1.87~MVAr. By comparison, the EMT cases in Table~\ref{tab:q_summary} settle within approximately 1.3~MVAr of the studied $\pm65.8$~MVAr PPC recovery magnitude. This quantitative accounting shows why a machine-side capability bound, together with the case-dependent network offset, need not reproduce the PPC-level recovery observed in EMT. The study does not rank all possible control parameters, identify a unique EMT-equivalent parameter set, or establish $K_c$ as the sole source of the EMT--RMS mismatch.

Taken together, the initialization and sensitivity results support a two-part diagnosis. First, initialization feasibility depends on compatibility between the initial control demand and the imposed hard limits, as in~\eqref{eq:init}. Second, after valid initialization with non-conflicting plant-controller bounds, post-disturbance recovery is shaped by both the active reactive bounds and plant-controller parameterization; sensitivity to $K_c$ becomes visible when the response moves off the bound. Thus, initialization consistency and recovery equivalence are distinct benchmarking requirements.

\section{Discussion and Limitations}
The numerical values reported here are not asserted to generalize to all IBRs. The limitation arises because the detailed EMT PPC structure, RMS control mode and parameters, initial reactive demand, network reactive losses, and active limiter state are project- and operating-point-dependent. The original lagging operating point violates the narrowed diagnostic \repca{} bound, whereas the added near-unity-PF point does not. Likewise, the measured $K_c$ sensitivity and saturation pattern are specific to the configured RMS control path and tested disturbances. What is intended to be transferable is the diagnostic procedure: establish parameter meaning and per-unit base, verify initialization compatibility, quantify the residual EMT--RMS difference, and then apply controlled RMS parameter variation together with limit-state checks to interpret the recovery behavior.

For this benchmark, the electrical-control capability values are retained in \reecc{}, while the wider \repca{} bounds are used as non-conflicting plant-controller bounds rather than interpreted as physical inverter capability. Conversely, the lower PPC recovery target should not be treated automatically as a hard physical capability limit. The observed $\Delta Q_{\mathrm{net}}$ variation from 25.71 to 36.55~MVAr also shows that a single fixed machine-side bound cannot generally impose the same POI recovery magnitude across the tested disturbance states. This supports retaining the POI-level recovery objective in a supervisory reference path rather than embedding it as a hard capability limit. An explicit supervisory-reference implementation is outside the demonstrated scope of the present generic RMS benchmark and is retained for future work.

\section{Conclusion}
This paper examined reactive-power recovery-limit interpretation in EMT--RMS benchmarking of a 200~MW anonymized IBR. Direct mapping of the $\pm0.2764$~pu PPC recovery target into \repca{} hard bounds was inconsistent at the original lagging operating point because $Q_0^{\rm cmd}=0.396273$~pu exceeded the imposed upper bound. With the same diagnostic bounds, a near-unity-PF operating point initialized successfully, confirming that the initialization conflict is conditional on operating point. Using the non-conflicting capability mapping removes the original hard-limit conflict, but the baseline comparison still shows 18.53--30.18~MVAr post-event POI reactive-power differences from PSCAD. The diagnostic mapping also initializes successfully for VDOWN, VUP, and leading LVRT yet retains 27.16--31.76~MVAr differences, demonstrating that valid initialization does not imply recovery equivalence.

A targeted $K_c$ sweep under matched benchmark initial conditions further shows why the RMS response can become insensitive to controller-parameter changes. At $K_c=0.05$ and 0.10, aggregate machine reactive output reaches the configured capability bounds; at $K_c=0.15$, VDOWN, VUP, and lagging HVRT move off those bounds and their settled POI reactive-power levels shift by 8.94, 18.00, and 3.17~MVAr, while leading LVRT remains at the positive bound. The results therefore show that hard-bound initialization feasibility, active-limit state, and post-event recovery equivalence must be evaluated as separate benchmarking checks rather than inferred from waveform overlay alone. Physical electrical capability and operational recovery targets should remain distinct control concepts during EMT--RMS benchmarking.

\balance

\end{document}